\documentclass{ws-procs11x85}

\makeindex
\begin{document}

\title{ASTROPARTICLE THEORY:\\
SOME NEW INSIGHTS INTO HIGH ENERGY COSMIC RAYS}

\author{ESTEBAN ROULET}

\address{CONICET, Centro At\'omico Bariloche, 8400, Argentina
\\E-mail: roulet@cab.cnea.gov.ar}


\twocolumn[\maketitle\abstract{ Some new developments obtained in the last 
few years concerning the propagation of high energy cosmic rays are discussed.
In particular, it is shown how the inclusion of drift effects in the transport 
diffusion equations leads naturally to an explanation for the knee, for the second 
knee and for the observed behavior of the composition and anisotropies between 
the knee and the ankle. It is shown that the trend towards a heavier composition 
above the knee has significant impact on the predicted neutrino fluxes
above $10^{14}$~eV. The effects of magnetic lensing on the cosmic rays with 
energies above the ankle are also discussed, analyzing the main features of 
the different regimes that appear between the diffusive behavior that takes 
place at 
lower energies and the regime of small deflections present at the highest ones.}]

\baselineskip=13.07pt
\section{Introduction}

Since their discovery in 1912, cosmic rays (CRs) were of great help for 
particle physics, providing a source of high energy particles for free, 
which only required the construction of detectors in order to observe different 
kinds of interesting phenomena.
In this way, positrons, muons, pions, kaons and hyperons were discovered 
in the period 1930--1950. However, when after the '50s man made accelerators 
reached energies beyond the GeV, particle physics moved back to the labs and 
cosmic ray research became focused on the study of the CRs themselves (rather 
than on the products of their interactions), trying to understand 
their origin, the mechanisms responsible for their acceleration and the way 
they propagate from the sources up to us.

There are only a few observable quantities  associated to the CR fluxes. 
These are the energy spectrum, the CR composition and the anisotropies in 
arrival directions, and it is through their study that the CR mysteries 
have to be unraveled.
For instance,  the fact that the 
spectrum is essentially a power law and not a thermal one is what led Fermi 
to suggest that the CR acceleration was a stochastic process. Also, looking 
at low energy cosmic rays, the study of 
the abundances of spallation products (like Li, Be and B, which are produced 
mainly by spallation of C, N and O) gives information about the amount of matter 
traversed by the CRs, while the abundances of unstable isotopes (e.g. ${}^{10}$Be)
gives information on the time spent by CRs in the Galaxy. On the other hand, the 
anisotropies observed on the low energy CRs arriving from the East and from the 
West gave indications that they were caused by the deflections produced by the 
geomagnetic field on the positively charged CRs.

Many puzzles are also associated with the observed properties of the CRs with 
very high energies, those above $10^{15}$~eV and up to the highest ones exceeding 
$10^{20}$~eV. In particular, we would like to know what causes the spectral 
changes observed, what is the origin of the observed anisotropies and the changes 
in composition as a function of energy 
and how CR sources would look like at ultrahigh energies. As 
will be discussed below, a crucial issue in this respect is to understand in 
detail the propagation of CRs through the magnetic fields present in the Galaxy, 
since this is essential to determine their properties when we finally observe them.

\section{The cosmic ray spectrum}

The differential flux of CRs changes by more than 30 orders of magnitude in the 
energy range from $10^9$~eV to $10^{20}$~eV, following essentially a power 
law $E^{-\alpha}$ which shows only some small but noticeable breaks in the 
power index $\alpha$.
One has indeed  $\alpha \simeq 2.6 \div 2.7$ at low energies, with a 
steepening to $\alpha\simeq 3$ at the so-called {\it knee}, which takes 
place at $E_k\simeq 3\times 10^{15}$~eV. A second steepening to $\alpha\simeq 3.3$,
referred to as the {\it second knee}, has been reported at $E_{sk}\simeq 
4\times 10^{17}$~eV, while a hardening to $\alpha\simeq 2.7$ takes place at 
$E_a\simeq 5\times 10^{18}$~eV, being known as the {\it ankle}.
 These general features 
are apparent in the data\cite{na00} shown in Fig.~1, 
 which displays the quantity $E^3\times {\rm d}\Phi/{\rm d}E$.
The observed spectrum is just a reflection of the original source spectra 
which have been reshaped by the energy dependence of  the confinement time 
of CRs in the Galaxy, since for increasing energies CRs escape more readily 
and hence their local density becomes less enhanced.
To understand this last effect we will then discuss in some detail
the transport of CRs through the magnetic fields permeating the Galaxy.
Beyond the ankle the CRs are no longer confined inside the Galaxy (and
most likely there are no galactic sources reaching such high  energies) so 
that a transition to an extragalactic component should be taking place there.

\begin{figure}
\center
\psfig{figure=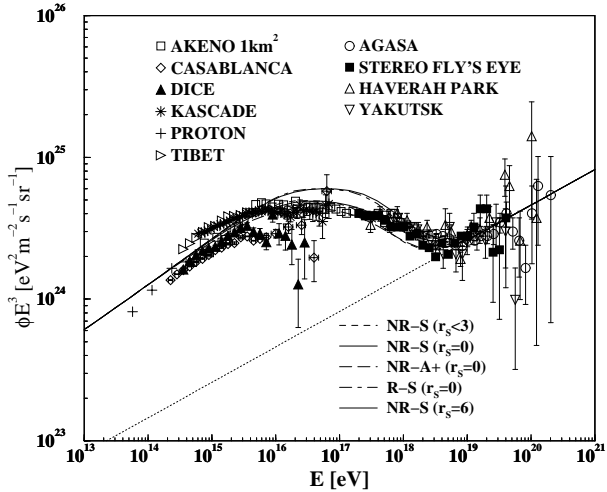,width=8.0truecm}
\caption{CR spectrum as measured by different experiments and predictions 
of the diffusion/drift scenario for different galactic models. The dotted
line is the assumed extragalactic flux.}
\label{crspectfit}
\end{figure}

\subsection{Cosmic ray diffusion and drift:}

Our Galaxy is believed to have a regular and a random magnetic field, 
both with local strength of a few $\mu$G. The regular component is aligned 
with the spiral arms, reversing its directions between consecutive arms and 
having a typical scale height of $\sim 1$~kpc. Also a more extended halo 
component is believed to exist, with typical scale height of a few 
kiloparsecs but its symmetry properties are less well established.
On the other hand, the random component has a spectrum not very 
different from a Kolmogorov one (as would result if its origin
is associated with the turbulence in the interstellar medium (ISM)), 
i.e. with a magnetic energy density satisfying d$E_r/{\rm d}k\propto k^{-5/3}$ 
in Fourier space, with a maximum scale of turbulence 
of order $L_{max}\simeq 100$~pc.

When particles of charge $Ze$ propagate across a regular field ${\bf B}_0$, they 
describe helical trajectories characterized by a pitch angle $\theta$, so 
that the component of the velocity parallel to ${\bf B}_0$ is $v_\parallel 
= c \cos\theta$, and a Larmor radius given by
\begin{equation}
r_L={pc\over ZeB_0}\simeq {E/Z\over 10^{15}\ {\rm eV}}{\mu 
{\rm G}\over B_0}{\rm pc}
\end{equation}
(the radius of the helical trajectory is $r_L\sin\theta$).
In the presence of the random component ${\bf B}_r$, the CRs will scatter 
off the magnetic field irregularities with associated scales of order 
$r_L$, changing their pitch angle but not their velocity. This will lead 
to a random walk and a diffusion along the magnetic field direction 
characterized by a diffusion coefficient
\begin{equation}
D_\parallel = {\langle \Delta x_\parallel^2\rangle\over 2\Delta t}=
{c\over 3}\lambda_\parallel,
\end{equation}
where $\lambda_\parallel$ is the pitch angle scattering length, which
depends on how much power there is in the magnetic field modes
at scales $\sim r_L$, i.e.
\begin{equation}
\left. \lambda_\parallel\propto {r_L\over {\rm d}E_r/{\rm d\ 
ln}k}\right|_{k=2\pi/r_L}.
\end{equation}
Hence, for a Kolmogorov spectrum one has $D_\parallel \propto E^{1/3}$, and
in general if the spectrum of the random magnetic field energy satisfies
d$E_r/{\rm d}k\propto k^{m-2}$, one has $D_\parallel\propto E^m$.

The diffusion orthogonal to the regular magnetic field direction
is typically much slower (unless the turbulence level is very high, 
in which case parallel and perpendicular motions become similar), and it  is
due to both pitch angle scattering and to the wandering of the magnetic 
field lines themselves, which drag with them the diffusing particles in 
the direction perpendicular to ${\bf B}_0$. Its evaluation has then to be 
done numerically, and recent results \cite{ca02} show that 
 for fixed levels of turbulence (i.e. for given 
values of $\sqrt{\langle {\bf B}_r^2\rangle}/B_0$) the
associated diffusion coefficient $D_\perp$ has a similar energy 
dependence as $D_\parallel$  as long as $r_L<L_{max}$.

\begin{figure}
\center
\psfig{figure=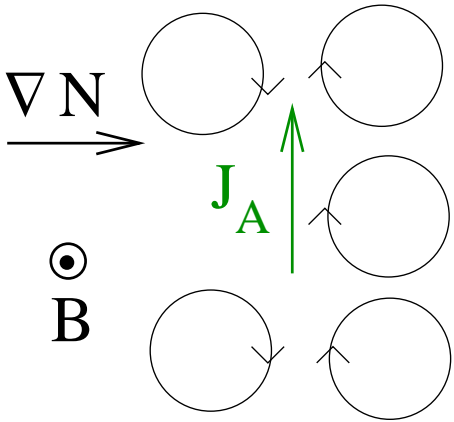,width=6.0truecm}
\caption{Illustration of the physical origin of the macroscopic drift.}
\label{gradrift}
\end{figure}

The third ingredient is the antisymmetric (or Hall) diffusion\cite{pt93}, which is 
associated to the drift of the CRs moving across the regular magnetic field. 
What is relevant here is the macroscopic drift associated to the gradient of 
the CR density, which leads to a current
\begin{equation}
J_A=D_A{{\bf B}_0\over B_0}\times \nabla N.
\end{equation}
Notice that this macroscopic current is orthogonal to both ${\bf B}_0$ and 
$\nabla N$, and its relation to the microscopic drift associated to gradients 
and curvature in ${\bf B}_0$ is subtle\cite{bu85}. 
In particular, the macroscopic drift is also present for a constant  ${\bf B}_0$, 
and its origin is illustrated in Fig.~2. The antisymmetric diffusion 
coefficient is just given by
\begin{equation}
D_A\simeq {r_L c\over 3}\propto E.
\end{equation}
The CR density distribution in the Galaxy is then determined by the equation
\begin{equation}
\nabla\cdot J_D=Q, \ {\rm with}\ J_{Di}=-D_{ij}\nabla_jN,
\label{diffeq}
\end{equation}
where $Q$ describes the distribution of sources and the diffusion tensor is 
(adopting here the $z$ axis along the direction of ${\bf B}_0$)
\begin{equation}
D_{ij}=\pmatrix{D_\perp & D_A & 0\cr
-D_A & D_\perp & 0 \cr
0 & 0 & D_\parallel}.
\end{equation}
To solve equation~(\ref{diffeq}) it 
is convenient to assume for simplicity that the 
Galaxy has cylindrical symmetry and that the magnetic field is azimuthal, 
what is not far from being true, in which case $D_\parallel$ plays no role and 
one has 
\begin{equation}
J_D\simeq -D_\perp \nabla N + D_A {{\bf B}_0\over B_0}\times \nabla N,
\end{equation}
so that the turbulent diffusive component is orthogonal to the 
isodensity contours while the drifts are parallel to them.

\begin{figure}
\center
\psfig{figure=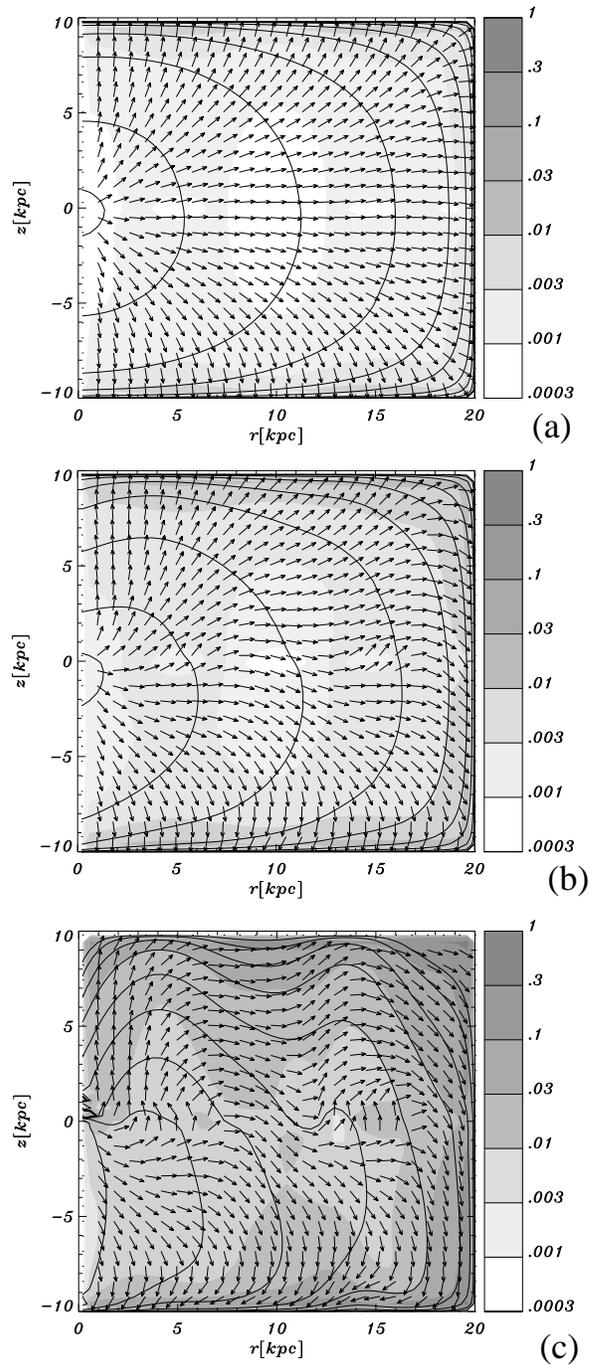,width=8.0truecm}
\caption{CR isodensity contours for $E/Z=10^{14},\ 3\times 10^{15}$ and 
$5\times 10^{17}$~eV (from top to bottom). The arrows indicate the directions 
of the diffusion currents while the shading the level of anisotropy.}
\label{galdriftv}
\end{figure}

Hence, adopting a realistic configuration of magnetic fields, obtaining 
from them the diffusion tensor everywhere in the Galaxy (using in particular 
fits to the numerical results obtained for $D_\perp$ for different turbulence 
levels\cite{ca02}), assuming a distribution of CR sources and then numerically solving the 
diffusion equations, one can obtain\cite{driftk,drifta} 
the CR density distribution in the Galaxy 
and the resulting diffusion currents.
This is exemplified in Fig.~3, which shows the isodensity contours
(every two solid contours correspond to a change in 
density by an order 
of magnitude) for a source of constant strength  in the inner 3~kpc
of the  galactic plane. The arrows indicate the direction of the
diffusion currents, with the component parallel to the contours
arising from the drifts. Notice that since the motion of a charged particle in a
magnetic field is determined just by its rigidity, the relevant
variable is the ratio $E/Z$. The three panels shown correspond to
$E/Z= 10^{14},\ 3\times 10^{15}$ and $5\times 10^{17}$~eV, and it is
clearly seen that for energies below $ZE_k$ the dominant effect
arises from the perpendicular diffusion, while for larger energies the
drifts are responsible for the dominant escape mechanism. The
transition between these two regimes is naturally understood from the
different energy dependence of the two diffusion coefficients
($D_\perp\propto E^{1/3}$ while $D_A\propto E$). Actually, both
coefficients become comparable, $D_\perp\simeq D_A$, at an energy of
the order of $ZE_k$, and this crossover naturally generates 
the observed break in the spectrum.

\subsection{The knee:}

If the source spectrum is taken as a power law, d$Q/{\rm d}E\propto
E^{-\alpha_s}$, it is seen from the diffusion equation that the
observed spectrum will be affected by the energy dependence of the
diffusion coefficients. This just reflects the fact that the CR
confinement time in the Galaxy is $\tau_e\propto D^{-1}$. Hence, below
the knee, where transverse turbulent diffusion  dominates, the
observed spectral index  will be $\alpha\simeq \alpha_s+1/3$, while in
the drift dominated regime one has instead $\alpha\simeq
\alpha_s+1$. Since this transition takes place at different energies
for CRs with different charges, one expects that first the proton
component will steepen its spectrum at an energy that will be
identified with $E_k$, and then the heavier components will suffer a
similar change in their spectrum but at energies $ZE_k$. This is
illustrated in Fig.~4, which shows the local density of the different 
CR components obtained\cite{driftk} in the scenario described in Fig.~3. The CR
composition is normalized with existing satellite measurements of
lower energies, typically below $10^{14}$~eV, and the individual
spectra of H, He, O and Fe are displayed, together with the total flux.

\begin{figure}
\center
\psfig{figure=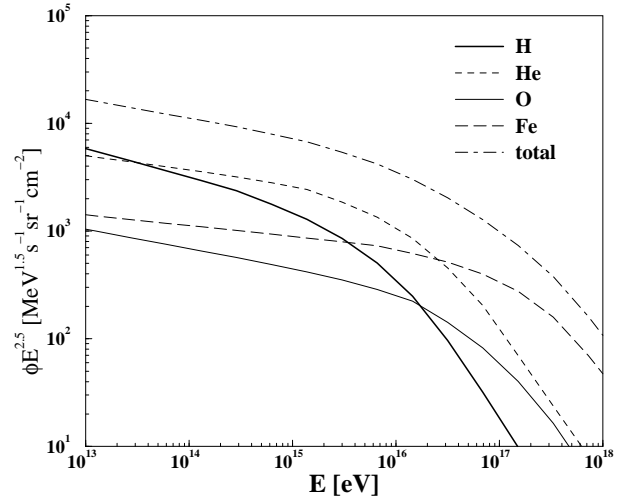,width=8.0truecm}
\caption{Predicted individual and total spectra in the diffusion/drift scenario.}
\label{compspec}
\end{figure}

It is seen that although each individual component changes its spectral 
index by $\Delta \alpha_i\simeq 2/3$, the total flux has a softer 
steepening, which is consistent with the measured change of 
$\Delta\alpha\simeq 0.3$.
Here the source spectrum is assumed to have a constant spectral index
$\alpha_s\simeq 2.3$ (which is consistent with expectations from first
order Fermi acceleration in strong shocks, which typically predict
$\alpha_s\simeq 2\div 2.4$) and the spectral change in the observed
spectrum just results from not ignoring the physical effects of the CR
drifts. Let us mention that the energy at which the break occurs
depends on several things, like the assumed distribution of sources,
the location in the Galaxy where the flux is measured, the symmetry of
the regular magnetic field, the level of turbulence and the maximum
scale of turbulence (scaling in particular as $L_{max}^{-1/2}$). It is
very reassuring that for the realistic models adopted it just falls in
the energy range where the knee is observed.

\subsection{The second knee and the ankle:}
Since the heaviest CRs having a significant abundance are the Fe
nuclei, once one considers energies  above $26\times E_k\simeq
10^{17}$~eV the light components will be already very suppressed and
hence the composition will be dominated  by these heavy nuclei. When 
drifts then start to affect this component, the total spectrum will
gradually  steepen to $\alpha\simeq \alpha_s+1\simeq 3.3$, which is
just the behavior observed at the so-called second-knee. This is
illustrated by the different lines depicted in Fig.~1, which show the
total spectrum obtained\cite{driftsk} with different magnetic field and source
models. In addition, the straight dotted line represents the
contribution from  extragalactic CRs, with an assumed spectrum
$\propto E^{-2.7}$ and normalized to fit observations above the ankle
(at energies beyond $5\times 10^{19}$~eV, the effects of the GZK
suppression by interactions with the CMB photons, not included here,
should also be accounted for). It is seen that all the main features of the
spectrum below the ankle find a natural explanation in terms of the CR
diffusion process.

\subsection{Composition and anisotropies:}

\begin{figure}
\center
\psfig{figure=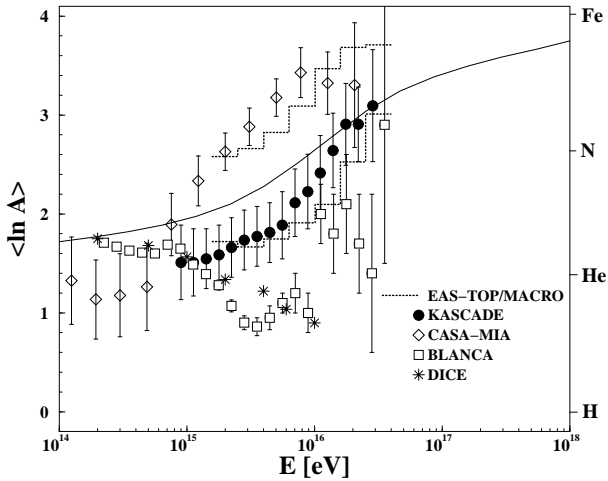,width=8.0truecm}
\caption{Measured composition and predictions (solid line) of the 
diffusion/drift scenario.}
\label{composit}
\end{figure}

As regards the composition, the behavior just described implies that
the CRs become increasingly heavy beyond the knee.  The predictions\cite{driftk}
for the average mass number ($\langle {\rm ln}A\rangle$) are displayed
in Fig.~5, together with the observational data on this quantity.
Although there was a large spread  on the measured values, the most
recent results from the KASCADE\cite{kascade} and the MACRO/EAS-TOP\cite{macro} 
Collaborations are in
quite good agreement between them and with the theoretical predictions
of the diffusion/drift scenario. Moreover, KASCADE has reported a
measurement of the spectra of four different mass groups (H, He, C, Fe) and they are
remarkably consistent with the predictions depicted in Fig.~4.  On the
other hand, the MACRO/EAS-TOP experiment has found that the change in
the spectral slope of the individual components is
$\Delta\alpha_i=0.7\pm 0.4$, consistent with the value of 2/3 induced
by the drifts.

Regarding the anisotropies, if the knee is actually due to a change in
the escape mechanism from the Galaxy (rather than a change in the
intrinsic source spectra), one expects to find a correlation between
the change in the behavior of the energy spectrum and of the CR
anisotropies. This is in fact what seems to be observed, and the
predicted\cite{drifta} anisotropies are in good agreement with the observational
results, as is apparent from Fig.~6.

\begin{figure}
\center
\psfig{figure=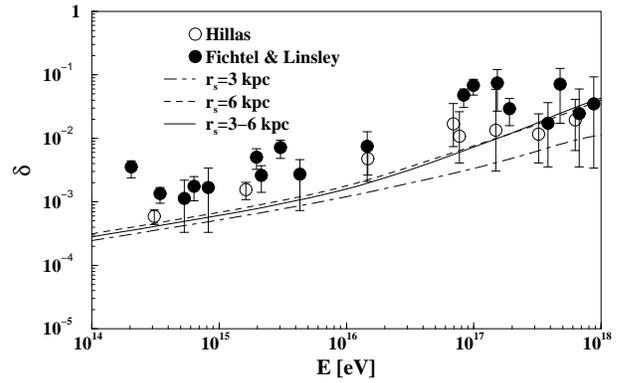,width=8.0truecm}
\caption{Measured anisotropies (amplitude of the first harmonic in 
right ascension)  and predictions of the diffusion/drift scenario 
for different assumed source distributions.}
\label{anisotropia}
\end{figure}

Let us mention that several possible mechanisms have been proposed 
in the past to explain the knee. 
For instance, that it could be associated to the heavy CR components
being photodisintegrated by interactions with optical or UV photons
inside the sources\cite{phdis}. 
This scenario would predict for instance that the
composition should become lighter above the knee, at odds with the
latest observations mentioned previously. It has also been suggested
that the knee could be related to a change in the efficiency of CR
acceleration in the sources. This is attractive since the traditional
acceleration scenario based on supernovae exploding in the ISM is
expected to lead to maximal energies of about $E/Z\sim 10^{14}\div
10^{15}$~eV. However, these scenarios (and also the one with
photodisintegrations) would have difficulties to account for the CR
anisotropies observed, for the origin of the second knee
and, in any case, since drift effects are anyhow present, they will
have to be taken into account and would lead to further suppressions beyond
those originating in the sources. Other acceleration mechanisms, such
as supernovae exploding  in the wind of their progenitors or  one-shot
acceleration in pulsars, may naturally reach larger values of $E/Z$,
so that the assumption that $\alpha_s$ remains constant beyond $E_k$
is quite plausible.

\section{Cosmic Ray induced neutrino fluxes}

When CRs interact  in the upper atmosphere producing particle cascades, a very
important product which results are the fluxes of muon and electron neutrinos, 
which arise mainly from the decays of charged pions and kaons. The study of
these atmospheric neutrinos has been of paramount importance in recent years,
allowing in particular to establish the existence of neutrino flavor
oscillations, and hence of neutrino masses and mixings. One may then say that
in this respect CRs have returned to be a source of particle beams of great
interest for particle physics. 

Also the interactions of galactic CRs with the gas present in the ISM is
expected to yield fluxes of high energy neutrinos from directions close to the
galactic plane (ISM $\nu$'s). Actually, the associated fluxes of high energy 
photons from $\pi^0$ decays  have already been observed by EGRET. In addition,
several galactic (supernova remnants, microquasars, ...) and extragalactic
(active galaxies, gamma ray bursts, ...) objects are also potential sources of
very high energy neutrinos, being the target of the ongoing and projected
neutrino observatories (Amanda, Baikal, Antares, Auger, Icecube, ...). Since
the flux sensitivities achieved are expected to improve by more than two
orders of magnitude in the next decade\cite{nufuture}, this will open a new window  to
explore the Universe by means of the high energy neutrinos. These messengers
have the great advantage of arriving undeflected and unattenuated from their
sources, containing then precious information about their production sites.

\begin{figure}
\center
\psfig{figure=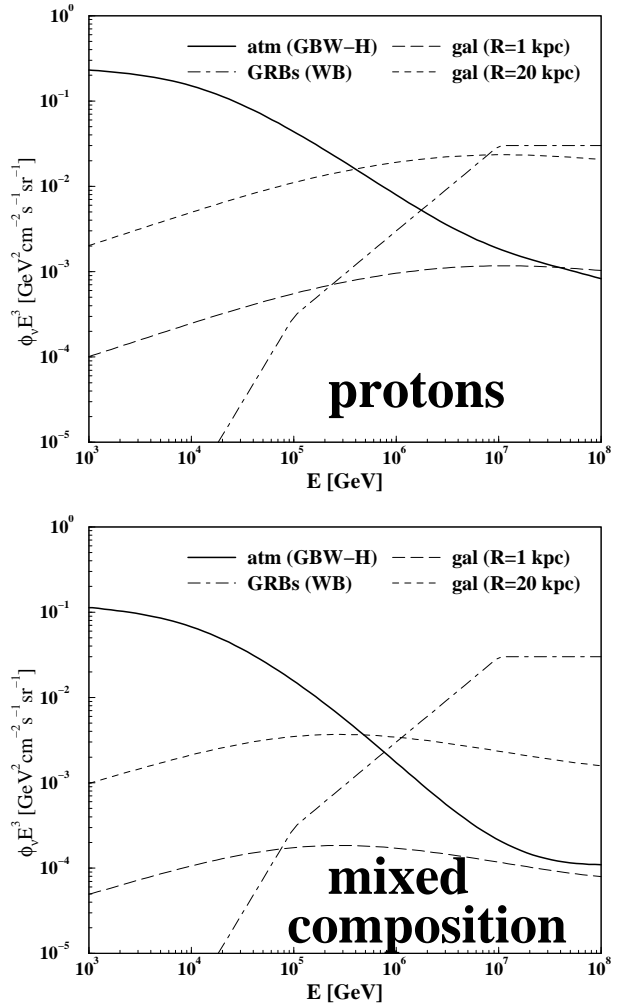,width=8.0truecm}
\caption{CR induced neutrino fluxes  under the assumption that the CR are 
only protons (top) or have a mixed composition (bottom), as results in 
rigidity dependent explanations of the knee. The solid line is the flux 
of atmospheric neutrinos, the dashed lines are the ISM $\nu$s from the 
direction of the galactic center (upper) or perpendicular to the galactic 
plane (lower line). Also indicated is the $\nu$ flux  from GRBs predicted 
by Waxman and Bahcall.}
\label{nuflux}
\end{figure}

CRs with energies beyond $E_k$ will produce atmospheric neutrinos   (and ISM
$\nu$s) with very high energies, typically $E_\nu>10^{14}$~eV. Atmospheric
neutrinos with these energies are particularly interesting because they are
expected to arise mainly from the decays of mesons containing charm
quarks, since these decay ``promptly'' after being produced while, due to the
relativistic time dilation, the pions and kaons have decay lengths much larger
than 10~km at these energies, and are hence stopped by the atmosphere before
they can decay, so that the neutrinos they produce have much lower
energies.

To evaluate the fluxes of prompt neutrinos is quite delicate, since
it requires\cite{charm} to take into account next to leading order processes in the charm
production cross section, which also turns out to depend on the behavior of
the parton distribution functions (PDF) at very low values ($x<10^{-4}$) of
the fraction of momentum carried by the partons, and these values are below
those that can be measured with present accelerators. It has been actually
suggested that the observations of very high energy atmospheric neutrinos
could be used to study the PDFs in the range $10^{-9}<x<10^{-4}$. Moreover, it
is important to know these fluxes in detail because they constitute  the main
background for the search of some astrophysical neutrino sources. 

To predict these fluxes it actually  turns out that one needs a detailed 
knowledge of the CR composition beyond the knee, since a heavy CR nucleus of
mass $A$ behaves just as a collection of $A$ nucleons of energy $E/A$. Hence,
if the CRs are heavy they will produce  fluxes of neutrinos  of much lower
energies, what combined with the steepness of the CR spectrum implies a much
reduced neutrino flux.  This is illustrated in Fig.~\ref{nuflux}, which shows
the neutrino fluxes predicted under the (standard) assumption that CRs
consist mainly of protons as well as the predictions adopting the composition
which results in models where the knee is a rigidity dependent effect (such as
the diffusion/drift scenario just discussed), and for which the composition
becomes heavier above the knee\cite{driftnus}.
Clearly this implies that the CR induced neutrino fluxes will be harder to
detect, but also that the background  for the search of other astrophysical
neutrino sources (such as the GRB prediction also shown in the figure) 
will be reduced.  

\section{Magnetic lensing}

\begin{figure}
\center
\psfig{figure=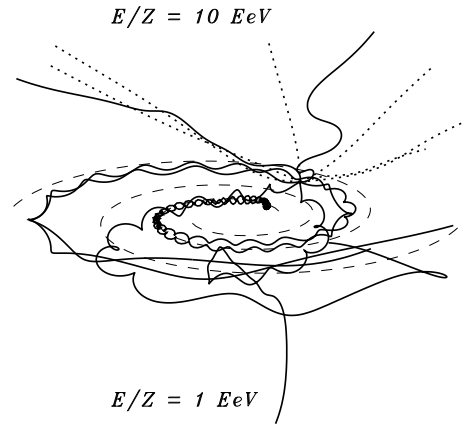,width=8.0truecm}
\caption{CR trajectories reaching the Earth for energies $E/Z=10^{18}$~eV 
(1~EeV, solid lines) 
and $E/Z=10^{19}$~eV (10~EeV, dotted lines). 
The dashed lines indicate the location of the galactic spiral arms.
}
\label{xyz}
\end{figure}

CRs have a diffusive propagation in the Galaxy as long as their Larmor radius
stays smaller than the maximum scale of the turbulence, i.e. for $E/Z$ below a
few times $10^{17}$~eV (see Eq.~(1)). 
It is only at higher energies that the trajectories
gradually straighten up and that it may become possible to attempt to do ``CR
astronomy'', i.e. to try to identify the images of individual CR sources. 
This is illustrated in Fig.~\ref{xyz}, which shows\cite{toes} several 
trajectories of CRs reaching the Earth. At energies $E/Z=10^{18}$~eV CRs 
are strongly deflected and there is no way in which the direction of arrival 
to the Earth can give information about the original source direction, while 
for $E/Z=10^{19}$~eV the deflections start to become manageable.

The transition between the diffusive behavior to the regime of small deflections
is actually very rich in new phenomena and several qualitatively different
regimes appear. 

\subsection{The regime of small deflections:}

Starting from the high energy end, we can expect that CRs arriving from
distant sources will suffer only small deflections, so that an image with a
slight (rigidity dependent) offset with respect to the original source
position should be observed. If the intervening magnetic field and the CR
charge were known, this effect could in principle be corrected in order to
infer the source coordinates\cite{st97,me98,reconstruct}. To have an idea about 
the deflections involved,
one can keep in mind that the deflection in the direction of the trajectory of
a CR  traversing a distance $L$ across a constant magnetic field $B$ is
\begin{equation}
\delta\simeq 3.2^\circ {10^{20}\ {\rm eV}\over E/Z}{L\over 3\ 
{\rm kpc}}{B\over 2\ \mu{\rm G}} .
\end{equation}
Hence, the regular galactic magnetic field 
can produce sizeable deflections even for
CRs with the highest energies observed, specially if their composition is
heavy. It has to be said however that almost nothing is known about the CR
composition above the ankle, and this is one of the issues that the future 
large statistics observatory AUGER is expected to clarify.  

Notice that although the random magnetic field has an rms strength similar to
the regular magnetic field strength, its effects on the overall CR deflections
is smaller, due to its short coherence length\footnote{For a Kolmogorov
  spectrum, one has $L_c\simeq L_{max}/5$.} $L_c$, which results in a deflection
growing only as $\sqrt{L/L_c}$, where $L/L_c$ is just the number of magnetic
``domains'' traversed. Indeed, the rms deflection it induces is
\begin{equation}
\delta_{rms}= 0.6^\circ {10^{20}\ {\rm eV}\over E/Z} {B_{rms}\over 
4\ \mu {\rm G}}\sqrt{L\over 3\ {\rm kpc}}\sqrt{L_c\over 50\ {\rm pc}}.
\end{equation}
The random component has however important effects on the lensing of CRs.
Let us also mention that extragalactic magnetic fields can further affect 
the CR trajectories\cite{si99,me98b,st01}. 
Their effects could be significant if they have sizeable strength and large 
coherence lengths (i.e. if 
$B\times \sqrt{L_c}>10^{-9} \ {\rm G}\sqrt{\rm Mpc}$).

\subsection{The lensing regime:}

When the deflections produced by the magnetic fields are such that the CRs can
arrive to the observer through paths which deviate from the straight one by
more than the scale of coherence of the magnetic field (e.g. a few kpc for the
regular field or the scale $L_c$ for the random component), it becomes
possible for the CRs to reach the observer  through different, uncorrelated,
paths. This just means that several different images of the same source will
appear. This phenomenon is quite similar to the well known gravitational
lensing phenomena, but here it is the magnetic deflection that replaces the
gravitational one. Contrary to this last one, the magnetic deflections are
energy dependent, so that the multiple images would look differently at
different energies. Anyhow, many magnetic lensing phenomena look as a function
of energy very similar  to some gravitational lensing phenomena as a function
of time when the relative motion of the lens is relevant, such as in the
microlensing events.

Besides producing multiple images, the magnetic field can also 
lead\cite{toes,signatures} to
focusing effects which magnify or demagnify the original source fluxes. In
particular, when new images appear at a certain energy $E_0$, they always
appear in pairs (which moreover have opposite parities, i.e. one of the images
will be inverted, something which in practice is of course unobservable
however), and they are magnified by a factor $A\propto (E_0-E)^{-1/2}$, which
diverges\footnote{The average magnification in a finite energy bin stays of
  course bounded.} 
for $E\to E_0$. This can lead to huge enhancements in the CR fluxes
in narrow ranges of energies.

\begin{figure}
\center
\psfig{figure=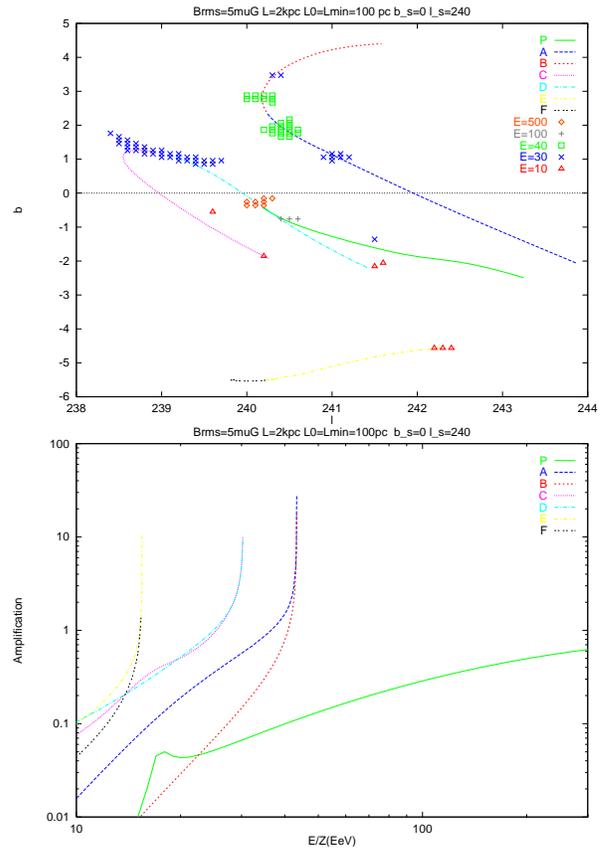,width=8.0truecm}
\caption{Images (top panel) of a CR source located near $(b,\ell)\simeq 
(0^\circ, 240^\circ)$. The different symbols correspond to the images 
obtained through a ray-shooting simulation (assuming an extended source) 
at energies 
$E/Z=500,$ 100, 40, 30 and $10\times 10^{18}$~eV. New secondary images 
appear in pairs below some critical energies, and for decreasing energies 
get displaced along the lines indicated. The bottom panel shows the 
magnification of the fluxes of the different images.}
\label{images}
\end{figure}

The multiple image formation  is illustrated\cite{turbulent} in Fig.~\ref{images},
 which shows how a particular source will look like at
different energies, and how  for decreasing energies new images are produced
and continuously displaced. The lensing effects for this same source are
 illustrated in the second panel, which shows the amplifications of
 the different images, clearly displaying the presence of lensing peaks
 when new images appear.  These peaks can lead to interesting observational
 effects, since they will lead to an enhanced number of events in a narrow
 angular window, corresponding to the direction where the new pair of images
 appear in the sky, which will also be clustered in energy (or, actually, in
 rigidity). It is remarkable that this clustering in rigidity seems to be
 already present\cite{turbulent} in the clusters of pairs and triplets of events with small
 angular separations observed\cite{doublets} at the highest energies ($E>4\times
 10^{19}$~eV). The increase in statistics that will be achieved with Auger will
 allow these predictions to be tested with higher significance.

\subsection{The scintillation regime:}

\begin{figure}
\center
\psfig{figure=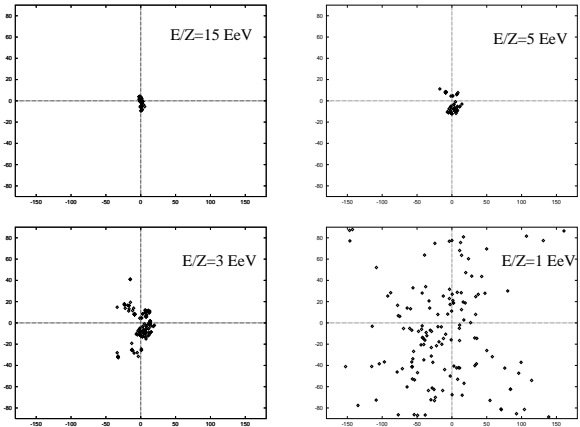,width=8.0truecm}
\caption{Images of a CR source in the scintillation regime.}
\label{panel}
\end{figure}

For typical galactic magnetic field models, multiple images of extragalactic
CR sources appear at energies $E/Z\sim {\rm few}\times 10^{19}$~eV. As smaller
energies are considered, one can show\cite{turbulent} 
that the number of secondary images
grows exponentially, increasing to about $10^2$ in a decade of energy
and this trend continues for lower energies. The lensing peaks associated to
the new images become increasingly narrow, and the result is that what is
observed is just a blurred image, consisting of a very large number of
secondary images, with an angular extent determined by $\delta_{rms}$ and with
an average magnification which approaches unity. This behavior is illustrated
in Fig.~\ref{panel}, which shows the results of a ray-shooting simulation
displaying the way  a given CR source will look like at different
energies in the regime in which a large number of images are present.

\begin{figure}
\center
\psfig{figure=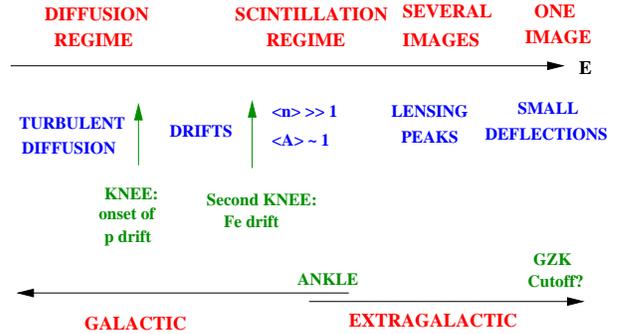,width=8.0truecm}
\caption{Sketch of the different CR propagation regimes.}
\label{regimes}
\end{figure}

This ``scintillation'' regime
is reminiscent to the twinkling of stars caused by the refractive effects of
the  atmospheric turbulence, although in the present case it is the galactic
magnetic turbulence the one responsible for the effects observed. This regime
is different from the diffusion one, since the deflections $\delta_{rms}$
need not be large. Hence, only some small random deflections in the direction
of the trajectories are being produced, but in order that the spatial
diffusion regime sets in one has to still go down in energy until $r_L<L_{max}$,
so that the resonant scattering leading to pitch angle scattering can take place.

A brief summary of the different propagation regimes discussed is shown in
Fig.~\ref{regimes}. 

It is interesting that we are at a time when many of these ideas will be
tested with more precise measurements of the CRs both in the region of the
knee (with e.g. KASCADE-GRANDE)  and at the highest energies (with AUGER), 
and this
will certainly help to unveil some of the many mysteries that CR still present
to us.

\section*{Acknowledgments}
This work is based on several collaborations done with J. Candia, 
L. Epele, D. Harari and S. Mollerach
It is partially supported by the Fundaci\'on Antorchas 
and the John Simon Guggenheim Foundation.

\end{document}